\documentclass[page-classic]{epl2} 

\usepackage{mathbbold}
\usepackage{amsmath}
\usepackage{array}
\usepackage{subeqnarray}
\usepackage{cases}
\usepackage[titletoc]{appendix}
\usepackage{graphicx}
\usepackage{subfigure}
\usepackage{dsfont}
\usepackage{amssymb}

\title{Lattice Boltzmann Model for the Volume-Averaged Navier-Stokes Equations}
\shorttitle{Lattice Boltzmann Model for The Volume-Averaged Navier-Stokes Equations} 
\author{Jinfeng Zhang\inst{1,2} \and Limin Wang\inst{1}\thanks{Corresponding author. E-mail: \email{lmwang@ipe.ac.cn}} \and Jie Ouyang\inst{2}\thanks{Corresponding author. E-mail: \email{jieouyang@nwpu.edu.cn}}}
\shortauthor{J.F. Zhang, L.M. Wang and J. Ouyang}
\institute{
  \inst{1} The EMMS Group, State Key Laboratory of Multiphase Complex Systems, Institute of Process Engineering, Chinese Academy of Sciences - Beijing 100190, China\\
  \inst{2} Department of Applied Mathematics, Northwestern Polytechnical University - Xi'an 710129, China}
\pacs{02.70.-c}{Computational techniques; simulations }
\pacs{05.10.-a}{Computational methods in statistical physics and nonlinear dynamics}
\pacs{47.11.Qr}{Lattice gas}
\abstract{A numerical method, based on the discrete lattice Boltzmann equation, is presented for solving the volume-averaged Navier-Stokes equations. With a modified equilibrium distribution and an additional forcing term, the volume-averaged Navier-Stokes equations can be recovered from the lattice Boltzmann equation in the limit of small Mach number by the Chapman-Enskog analysis and Taylor expansion. Due to its advantages such as explicit solver and inherent parallelism, the method appears to be more competitive with traditional numerical techniques. Numerical simulations show that the proposed model can accurately reproduce both the linear and nonlinear drag effects of porosity in the fluid flow through porous media.
}

\begin{document}

\maketitle
\section{Introduction}
The lattice Boltzmann method (LBM), which is developed from the lattice gas automata (LGA)\cite{HPP,FHP,Humieres1986,Succi1989,higuera1989a,higuera1989b,succi2001lattice} and considered as a powerful discrete scheme of the continuum Boltzmann equation\cite{he1997theory}, has attracted considerable attention in modeling hydrodynamics of flow systems. After more than two decades of development, LBM has been successfully applied in the simulation of fluid flow\cite{Koelman1991,qian1992,chen1998lattice,aidun2010lattice}.

By the Chapman-Enskog analysis and Taylor expansion\cite{qian1993,sterling1996,He1997,Qian1998}, the standard lattice Boltzmann equation (LBE) can recover the macroscopic Navier-Stokes equations that are usually used to describe simple or single-phase flow. When it comes to both the flow through porous media at representative elementary volume scale and the multiphase flow based on the idea of interpenetrating continua (namely multi-fluid model), the volume-averaged Navier-Stokes equations\cite{kuipers1992,VANS} are of opportune description. To date, two algorithms, namely the interphase slip algorithm (IPSA)\cite{spalding1980numerical} and the implicit multi-field (IMF) method \cite{harlow1975numerical}, are mainly applied to solve such equations. Additionally, when the coupling of phases in multiphase flows is very strong for the multi-fluid model, the interphase coupling algorithms, such as the partial elimination algorithm (PEA)\cite{spalding1980numerical} and the simultaneous solution of non-linearly coupled equations (SINCE)\cite{karema1999efficiency} etc, are applied to avoid the divergence of iterative sequential solvers. However, almost all the mentioned algorithms are implicit/semi-implicit schemes, which suffer from relatively lower scalability and parallel efficiency.

LBM is a promising method for large-scale fast simulation of multiphase flow problems due to its advantages of explicit solver, simple coding and natural parallelism\cite{chen1998lattice,aidun2010lattice}. In the literature, various LBE formulations for modeling the volume-averaged Navier-Stokes equations have been reported\cite{guo2002lattice,wang2005two,sankaranarayanan2008lattice,sungkorn2012simulations,eggels1995numerical,wang2013lattice}. Guo et al.\cite{guo2002lattice} put forward a generalized lattice Boltzmann model for isothermal incompressible flows in porous media. Wang et al.\cite{wang2005two} reformulated the lattice Boltzmann equation as a dimensionless one, and added a modified pressure term to recover the volume-averaged Navier-Stokes equations through the Chapman-Enskog analysis. Sankarananrayanan et al.\cite{sankaranarayanan2008lattice} modified the equilibrium distribution by introducing the volume fraction and amended the continuous Boltzmann equation for considering the temperature of particle to ensure the mass and momentum conservation in gas-solid flow. Unfortunately, the macroscopic equations derived from all the above-mentioned LBEs are not equivalent to the full volume-averaged Navier-Stokes equations. Sungkorn and Derksen\cite{sungkorn2012simulations} reformulated the volume-averaged Navier-Stokes equations by putting all the volume fraction terms to the right hand side (RHS) as a source term or forcing term, and solved the reformulated equations by the FCHC method\cite{eggels1995numerical}. Recently, Wang et al.\cite{wang2013lattice} proposed a modified LBE with a reasonable consideration for the effect of both the local solid volume fraction and the local relative velocity between particles and fluid, while the corresponding macroscopic equations of the modified LBE were not established. Therefore, a rigorous theoretical derivation from the LBE to the volume-averaged Navier-Stokes equations is highly anticipated.

\section{Lattice Boltzmann model}
The lattice Boltzmann equation with BGK approximation can be expressed as

\begin{equation}
\label{e.BGK}
{f_{i}}({\bf{x}} + {{\bf{e}}_i}\delta t,t + \delta t) - {f_{i}}({\bf{x}},t) = -\frac{1}{\tau }[{f_{i}}({\bf{x}},t) - f_{i}^{\mathrm{eq}}({\bf{x}},t)],
\end{equation}
where ${f_{i}}({\bf{x}},t)$ is the single particle distribution function indicating the probability of finding a particle with velocity ${\bf e}_i$ at position $\bf{x}$ and time $t$, $\delta t$ is the discrete time step, $\tau$ is the relaxation time, and $f_{i}^{\mathrm{eq}}({\bf{x}},t)$ is the equilibrium distribution function. It is well known that the lattice Bhatnagar-Gross-Krook (LBGK) model is one of the most popular LB models and can recover the Navier-Stokes equations\cite{qian1992} given as follows:

\begin{subequations}  \label{e.NS}
\begin{align}
&\frac{\partial }{{\partial t}}\rho  + \nabla  \cdot (\rho {\bf{u}}) = 0,\label{e.NSA} \\
\frac{\partial }{{\partial t}}(\rho {\bf{u}}) + \nabla  \cdot (\rho {\bf{uu}}&) =  -  \nabla p + \nu \nabla  \cdot \{ \rho [\nabla {\bf{u}} + {(\nabla {\bf{u}})^{\rm{T}}}]\},\label{e.NSB}
\end{align}
\end{subequations}
where ${\rho }$ is the fluid density, ${\bf{u}}$ is the fluid velocity, $\nu $ is the kinematic viscosity, $p$ is the pressure.

In order to find a model for the volume-averaged Navier-Stokes equations, we begin with comparing the volume-averaged Navier-Stokes equations with the standard Navier-Stokes equations. To make the illustration more clear, we neglect the forcing term in the derivation. The volume-averaged Navier-Stokes equations\cite{kuipers1992,VANS} are extensively used to describe flow through porous media, multiphase flow and other complex flows, and can be expressed as follows:

\begin{subequations}  \label{e.VNS1}
\begin{align}
 &\frac{\partial }{{\partial t}}(\varepsilon \rho ) + \nabla  \cdot (\varepsilon \rho {\bf{u}}) = 0,\label{e.VNS1A} \\
 \frac{\partial }{{\partial t}}(\varepsilon \rho {\bf{u}}) + \nabla  \cdot (\varepsilon \rho &{\bf{uu}}) =  - \varepsilon \nabla p + \nu \nabla  \cdot \{\varepsilon \rho [\nabla {\bf{u}} + {(\nabla {\bf{u}})^{\rm{T}}}]\},           \label{e.VNS1B}
\end{align}
\end{subequations}
where $\varepsilon$ represents the volume fraction in multiphase flow or local porosity in porous media flow.

Comparing eqs.~(\ref{e.NS}) and  eqs.~(\ref{e.VNS1}), it is readily found that there exists a variable $\varepsilon$ before $\rho$ in the volume-averaged Navier-Stokes equations. Therefore, it is natural for us to introduce $\varepsilon$ before $\rho$ in the equilibrium equation, and we can get

\begin{equation}
\label{e.d2q9eq}
f_{i}^{\mathrm{eq}}({\bf{x}},t) = {w_i}{\varepsilon }{\rho }[1 + \frac{{{\bf{e}} \cdot {\bf{u}}}}{{c_s^2}} + \frac{{{{({\bf{e}} \cdot {\bf{u}})}^2}}}{{2c_s^4}} - \frac{{\bf{u}} \cdot {\bf{u}}}{{2c_s^2}}],
\end{equation}
where $w_i$ is the weight and $c_s$ is the sound speed.

By applying the Chapman-Enskog analysis and Taylor expansion, and omitting the high-order terms, the LBGK model with the above equilibrium equation can recover the following equations

\begin{subequations}  \label{e.VNS2}
\begin{align}
& \qquad\qquad \frac{\partial }{{\partial t}}(\varepsilon \rho ) + \nabla  \cdot (\varepsilon \rho {\bf{u}}) = 0,\label{e.VNS2A} \\
 \begin{split}
 \frac{\partial }{{\partial t}}(\varepsilon \rho {\bf{u}}) + \nabla  \cdot (\varepsilon \rho {\bf{uu}}) = & - \nabla(\varepsilon p)\\& + \nu \nabla  \cdot \{\varepsilon \rho [\nabla {\bf{u}} + {(\nabla {\bf{u}})^{\rm{T}}}]\},\label{e.VNS2B}\\
 \end{split}
\end{align}
\end{subequations}
where $p=c_s^2\rho$, and $\nu=c_s^2(\tau-0.5)\delta t$. We can see that the pressure term $-\nabla(\varepsilon p)$ is different from $-\varepsilon \nabla p$ in the volume-averaged Navier-Stokes equations. Furthermore, from the vector identity concerning the divergence of a scalar times a scalar,

\begin{equation}
 - \varepsilon \nabla p=p\nabla \varepsilon -\nabla(\varepsilon p),
\end{equation}
we can add a discrete term of $p\nabla \varepsilon$ (denoted by $P_i$) to the discrete Boltzmann equation with BGK approximation. Therefore, the proposed model can be expressed as

\begin{equation}
\label{e.vlbe}
\begin{split}
{f_{i}}({\bf{x}} + {{\bf{e}}_i}\delta t,t + \delta t) - {f_{i}}({\bf{x}},t) =&-\frac{1}{\tau }[{f_{i}}({\bf{x}},t) - f_{i}^{eq}({\bf{x}},t)]\\&+ \delta t P_i,
\end{split}
\end{equation}
where $f_{i}^{\mathrm{eq}}$ is given as eq.~(\ref{e.d2q9eq}), and the additional term $P_i$ can be written as\cite{force2002discrete}

\begin{equation}
\label{e.pi}
P_i = (1 - {\frac{1}{2\tau }}){w_i}({\frac{{{\bf{e}}_i} - {\bf{u}}}{c_s^2}} + {\frac{{{\bf{e}}_i} \cdot {\bf{u}}} {c_s^4}}{{\bf{e}}_i})\cdot p\nabla \varepsilon.
\end{equation}
As a result, the macroscopic equations corresponding to eq.~(\ref{e.vlbe}) are eqs.~(\ref{e.VNS1}).

\section{Chapman-Enskog analysis}
In the following, we would derive the corresponding macroscopic equations from the proposed lattice Boltzmann equations through the Chapman-Enskog analysis. With the equilibrium distribution function $f_{i}^{eq}$, as defined in eq.~(\ref{e.d2q9eq}), we get the following moment equations

\begin{subequations}
\begin{align}
&\sum\limits_i {f_{i}{^{\mathrm{eq}}}}  = {\varepsilon }{\rho },\\
\sum\limits_i& {{{\bf{e}}_i}f_{i}{^{\mathrm{eq}}}}  = {\varepsilon }{\rho }{\bf{u}},\\
\sum\limits_i {{{\bf{e}}_i}{{\bf{e}}_i}f_{i}^{\mathrm{eq}}}&  = c_s^2{\varepsilon }{\rho } + {\varepsilon }{\rho }{u_\alpha }{u_\beta },\\
\sum\limits_i {{{\bf{e}}_i}{{\bf{e}}_i}{{\bf{e}}_i}f_{i}^{\mathrm{eq}}} = c_s^2{\varepsilon }&{\rho }({\delta _{\alpha \beta }}{u_\gamma } + {\delta _{\alpha \beta }}{u_\gamma } + {\delta _{\beta \gamma }}{u_\alpha }).
\end{align}
\end{subequations}
With the additional term $P_i$, it is noted that

\begin{subequations}
\begin{align}
&\sum\limits_i {P_i}  = 0,\\
\sum\limits_i{{{\bf{e}}_i}P_i}& = (1 - {\frac{1}{2\tau }})c_s^2{\rho }\nabla {\varepsilon },\\
\sum\limits_i {{{\bf{e}}_i}{{\bf{e}}_i}P_i}= (1 - &{\frac{1}{ {2\tau }}})({\bf{u}}c_s^2{\rho }\nabla {\varepsilon } + c_s^2{\rho }\nabla {\varepsilon }{\bf{u}}).
\end{align}
\end{subequations}

Based on the Chapman-Enskog analysis and Taylor expansion, the following expressions are given

\begin{equation}
{f_{i}} = f_{i}^{(0)} + \lambda f_{i}^{(1)} + {\lambda ^2}f_{i}^{(2)}
\end{equation}
\begin{equation}
\frac{\partial }{{\partial t}} = \lambda \frac{\partial }{{\partial {t_1}}} + {\lambda ^2}\frac{\partial }{{\partial {t_2}}}
 \end{equation}
\begin{equation}
\nabla  = \lambda {\nabla _1}
\end{equation}
\begin{equation}
\begin{split}
{f_{i}}({\bf{x}} + {{\bf{e}}_i}\delta t,t + \delta t) = &{f_{i}}({\bf{x}},t) +\delta t(\frac{\partial }{{\partial t}} + {{\bf{e}}_i}\nabla ){f_{i}}({\bf{x}},t) \\
&+ \frac{{{\delta t}^2}}{2}{(\frac{\partial }{{\partial t}} + {{\bf{e}}_i}\nabla )^2}{f_{i}}({\bf{x}},t)
\end{split}
\end{equation}
where $\lambda$ is an expansion parameter, which is proportional to the ratio of the lattice spacing to a characteristic macroscopic length.

Applying the above expressions to eq.~(\ref{e.vlbe}), we can obtain the following equations in the consecutive order of the parameter $\lambda$:

\begin{widetext}
\begin{eqnarray}
\label{e.0}{\rm O}({\lambda ^0}):& &  f_{i}^{(0)} = f_{i}^{\mathrm{eq}} ,\\
\label{e.1}{\rm O}({\lambda ^1}):& & (\frac{\partial }{{\partial {t_1}}} + {{\bf{e}}_i}\cdot{\nabla _1})f_{i}^{(0)} =  - \frac{{f_{i}^{(1)}}}{{\tau \delta t}}+\frac{P_{i}}{\lambda},\\
\label{e.2}{\rm O}({\lambda ^2}):& & \frac{{\partial f_{i}^{(0)}}}{{\partial {t_2}}}                                                                                                              + (1 - {\frac{1}{2\tau }})(\frac{\partial }{{\partial {t_1}}} + {{\bf{e}}_i}\cdot{\nabla _1})f_{i}^{(1)} =- \frac{{f_i^{(2)}}}{{\tau \delta t}}- {\frac{\delta t}{2\lambda}}(\frac{\partial }{{\partial {t_1}}} + {{\bf{e}}_i}\cdot{\nabla _1})P_{i}.
\end{eqnarray}
\end{widetext}

\mbox{\textit{see eq.~\eqref{e.0},eq.~\eqref{e.1} and eq.~\eqref{e.2}.}} 

{The zero-order velocity moment of eq.~(\ref{e.1}) is

\begin{equation}
\sum\limits_i {(\frac{\partial }{{\partial {t_1}}} + {{\bf{e}}_i} \cdot {\nabla _1})f_{i}^{(0)}}  = -\sum\limits_i { \frac{{f_{i}^{(1)}}}{{\tau \delta t}}}+ \sum\limits_i {\frac{P_{i}}{\lambda} },
\end{equation}
from which we can get

\begin{equation}
\label{e.11vmoment}
\frac{\partial }{{\partial {t_1}}}({\varepsilon }{\rho }) + {\nabla _1} \cdot ({\varepsilon }{\rho }{{\bf{u}}}) = 0.
\end{equation}
The first-order velocity moment of eq.~(\ref{e.1}) is

\begin{equation}
 \sum\limits_i {{{\bf{e}}_i}(\frac{\partial }{{\partial {t_1}}} + {{\bf{e}}_i}\cdot{\nabla _1})f_{i}^{(0)}}  =  - \sum\limits_i {\frac{{{\bf{e}}_i}{f_{i}^{(1)}}}{{\tau \delta t}}} + \sum\limits_i {\frac{{{\bf{e}}_i}P_{i}}{\lambda} }  ,
\end{equation}
then, we can obtain

\begin{equation}
\label{e.12vmoment}
 \frac{\partial }{{\partial {t_1}}}({\varepsilon }{\rho }{{\bf{u}}}) + {\nabla _1} \cdot (c_s^2{\varepsilon }{\rho } + {\varepsilon }{\rho }{u_{\alpha}}{u_{\beta}})=\frac{ c_s^2{\rho }\nabla   {\varepsilon }}{\lambda}.
\end{equation}
And the zero-order velocity moment of eq.~(\ref{e.2}) is

\begin{equation}
  \begin{split}
 &   \sum\limits_i {\frac{{\partial f_{i}^{(0)}}}{{\partial {t_2}}}}  + (1 - {\frac{1}{2\tau }})\sum\limits_i {(\frac{\partial }{{\partial {t_1}}} + {{\bf{e}}_i}\cdot{\nabla _1})f_{i}^{(1)}}\\
  &=  - \sum\limits_i {\frac{f_i^{(2)}}{{\tau \Delta t}}} - \sum\limits_i {\frac{\delta t} { 2\lambda }(\frac{\partial }{{\partial {t_1}}} + {{\bf{e}}_i}\cdot{\nabla _1})P_{i}} ,
  \end{split}
\end{equation}
we can readily get

\begin{equation}
\label{e.21vmoment}
\frac{{\partial( {\varepsilon }{\rho })}}{{\partial {t_2}}}{\rm{ = }}0.
\end{equation}

Since the time derivative term can be expressed as

\begin{equation}
\label{e.timederivative}
\begin{split}
&\frac{\partial }{{\partial {t_1}}}({\varepsilon }{\rho }{u_{\alpha}}{u_{\beta}})\\&=  - {u_{\alpha}}{u_{\beta}}\frac{\partial }{{\partial {t_1}}}({\varepsilon }{\rho }) + {u_{\alpha}}\frac{\partial }{{\partial {t_1}}}({\varepsilon }{\rho }{u_{\beta}}) + {u_{\beta}}\frac{\partial }{{\partial {t_1}}}({\varepsilon }{\rho }{u_{\alpha}}) \\
&  =  \frac{1}{\lambda}({u_{\alpha}}c_s^2\rho\nabla \varepsilon +{u_{\beta}}c_s^2\rho\nabla \varepsilon) - ({u_{\alpha}} + {u_{\beta}}){\nabla _1} (c_s^2{\varepsilon }{\rho })\\
&\quad- {\nabla _1} \cdot ({\varepsilon }{\rho }{u_{\alpha}}{u_{\beta}}{{\bf{u}}}),
 \end{split}
\end{equation}
the momentum flux tensor can be simplified as

\begin{equation}
\label{e.fluxtensor}
\begin{split}
&\sum\limits_i {{{\bf{e}}_i}{{\bf{e}}_i}f_{i}^{(1)}}\\
&= \tau \Delta t({u_{\alpha}} + {u_{\beta}}){\nabla _1} ({\varepsilon }{\rho }{u_{\alpha}}{u_{\beta}})\\
&\quad - \tau \Delta t c_s^2{\varepsilon }{\rho }{\nabla _1}\cdot({\delta _{\alpha\gamma}}{u_\beta} + {\delta _{\gamma\beta}}{u_\alpha})\\
&\quad- {\frac{\Delta t}{2\lambda }}({\bf{u}}{c_s^2\rho\nabla \varepsilon} + c_s^2{\rho }\nabla   {\varepsilon }{\bf{u}})\\
&\quad - \tau \Delta t[{u_{\alpha}}{u_{\beta}}{\nabla _1} \cdot ({\varepsilon }{\rho }{{\bf{u}}})] .\\
\end{split}
\end{equation}
Once we have substituted eq.~(\ref{e.fluxtensor}) into the first-order velocity moment of eq.~(\ref{e.2}), we can obtain

\begin{equation}
\label{e.22vmoment}
\begin{split}
 &\frac{\partial}{{\partial {t_2}}} ({\varepsilon }{\rho }{\bf{u}}) + (1 - {\frac{1}{2\tau }}) \frac{\partial }{{\partial {t_1}}}( - {\frac{1}{2\lambda}}\Delta t{c_s^2\rho\nabla \varepsilon})\\
&- (1 - {\frac{1}{2\tau }}){\nabla _1}\cdot[\tau \Delta tc_s^2{\varepsilon }{\rho }{\nabla _1}\cdot({\delta _{\alpha\gamma}}{u_\beta} + {\delta _{\gamma\beta}}{u_\alpha})] \\
 & + (1 - {\frac{1}{2\tau }}){\nabla _1}\cdot[\tau \Delta t({u_{\alpha}} + {u_{\beta}}){\nabla _1} \cdot ({\varepsilon }{\rho }{u_{\alpha}}{u_{\beta}})] \\
&- (1 - {\frac{1}{2\tau }}){\nabla _1}\cdot\{\tau \Delta t[{u_{\alpha}}{u_{\beta}}{\nabla _1} \cdot ({\varepsilon }{\rho }{{\bf{u}}})]\} \\
 & =
  - {\frac{\Delta t}{2\lambda}} \frac{\partial }{{\partial {t_1}}}[(1 - {\frac{1}{2\tau }})c_s^2\rho\nabla \varepsilon]. \\
\end{split}
\end{equation}

In the above work, four macroscopic equations that are eq.~(\ref{e.11vmoment}), eq.~(\ref{e.12vmoment}), eq.~(\ref{e.21vmoment}) and eq.~(\ref{e.22vmoment}) have been obtained. Combining these equations, we have

\begin{subequations}  \label{e.VNS4}
\begin{align}
 &\qquad\qquad\frac{\partial }{{\partial t}}({\varepsilon }{\rho }) + \nabla  \cdot ({\varepsilon }{\rho }{{\bf{u}}}) = 0,\label{e.VNS4A} \\
\begin{split}
&\frac{\partial }{{\partial t}}({\varepsilon }{\rho }{{\bf{u}}}) + \nabla \cdot ({\varepsilon }{\rho }{u_{\alpha}}{u_{\beta}})  =  - {\varepsilon }\nabla   (c_s^2{\rho })\\
&\quad + \tau \Delta tc_s^2(1 - {\frac{1}{2\tau }})\nabla\cdot [{\varepsilon }{\rho }\nabla\cdot ({\delta _{\alpha\gamma}}{u_\beta} + {\delta _{\gamma\beta}}{u_\alpha})] \\
& \quad+ (1 - {\frac{1}{2\tau }})\nabla \cdot\{\tau \Delta t[{u_{\alpha}}{u_{\beta}}\nabla  \cdot ({\varepsilon }{\rho }{{\bf{u}}})]\} \\
&\quad- (1 - {\frac{1}{2\tau }})\nabla \cdot[\tau \Delta t({u_{\alpha}} + {u_{\beta}})\nabla  \cdot ({\varepsilon }{\rho }{u_{\alpha}}{u_{\beta}})]. \\
\end{split}           \label{e.VNS4B}
\end{align}
\end{subequations}

Omitting the high-order terms which are the divergents term representing the compressibility, we get eq.~(\ref{e.VNS1}),
where $p=c_s^2\rho$, $\nu=c_s^2(\tau-0.5)\delta t$. As a consequence, eq.~(\ref{e.vlbe}) can recover the volume-averaged Navier-Stokes equations without forcing term. An additional discrete forcing term should be added when recovering the volume-averaged Navier-Stokes equations with a forcing term\cite{kuipers1992,VANS}, namely

\begin{subequations}  \label{e.VNS3}
\begin{align}
 &\frac{\partial }{{\partial t}}(\varepsilon \rho ) + \nabla  \cdot (\varepsilon \rho {\bf{u}}) = 0,\label{e.VNS3A} \\
\begin{split}
\frac{\partial }{{\partial t}}(\varepsilon \rho {\bf{u}}) &+ \nabla  \cdot (\varepsilon \rho{\bf{uu}}) =  - \varepsilon \nabla p
\\&+ \nu \nabla  \cdot \{\varepsilon \rho [\nabla {\bf{u}}+ {(\nabla {\bf{u}})^{\rm{T}}}]\}+\bf{F},
\end{split}  \label{e.VNS3B}
\end{align}
\end{subequations}
Therefore, the final lattice Boltzmann equation for the volume-averaged Navier-Stokes equations with a forcing term can be expressed as:

\begin{equation}
\begin{split}
{f_{i}}({\bf{x}} + {{\bf{e}}_i}\delta t,t + \delta t) - {f_{i}}({\bf{x}},t) =&-\frac{1}{\tau }[{f_{i}}({\bf{x}},t) - f_{i}^{\mathrm{eq}}({\bf{x}},t)]\\&+ \delta t P_i+ \delta t F_i,
\end{split}
\end{equation}
where $P_i$ is given as eq.~(\ref{e.pi}),  $F_i$ is the discrete forcing term\cite{force2002discrete}, and is given as

\begin{equation}
\label{e.Fi}
F_i = (1 - \frac{1}{2\tau }){w_i}({\frac{{{\bf{e}}_i} - {\bf{u}}}{c_s^2}} + {\frac{{{\bf{e}}_i} \cdot {\bf{u}}}
{c_s^4}}{{\bf{e}}_i})\cdot\bf{F},
\end{equation}
The macroscopic values of fluid density and velocity are obtained from the moments of the particle distribution function

\begin{equation}
\label{e.0moment}
{\varepsilon }{\rho } = \sum\limits_i {{f_i}},
\end{equation}

\begin{equation}
\label{e.1moment}
{\varepsilon }{\rho }{\bf{u}}  = \sum\limits_i {{{\bf{e}}_i}{f_i}} + {\frac{1}{2}}\delta tc_s^2\rho\nabla \varepsilon + {\frac{1}{2}}\delta t\bf{F}.
\end{equation}

\section{Examples and discussions}
To validate the proposed lattice Boltzmann model, a two-dimensional Couette flow through porous media with porosity $\varepsilon$ (fig.~\ref{fig.1}) is simulated. The Couette flow is through porous media of porosity $\varepsilon$ between two plates separated by a distance $H$. The flow is driven by the upper plate with a constant velocity $u_0$ along the $x$ direction. A periodic boundary condition is applied in the $x$ direction. A no-slip boundary condition\cite{wang2006cpc} and a velocity boundary condition\cite{zouhe1997boundary} are imposed to the bottom and upper plates, respectively.

 The flow suffers resistance through porous media given as\cite{guo2002lattice}

\begin{equation}
 {\bf{F}} =  - \frac{{{\varepsilon ^2}\nu }}{K}{\bf{u}} - \frac{{{\varepsilon ^3}{F_\varepsilon }}}{{\sqrt K }}\left| {\bf{u}} \right|{\bf{u}}.
\end{equation}
Here, $\nu$ is the kinematic viscosity, $K$ and $F_\varepsilon$ represent the permeability and geometric function. Based on Ergun's experimental investigations\cite{Ergun}, $K$ and $F_\varepsilon$ can be expressed as\cite{vafai1984}

\begin{equation}
K=\frac{{{\varepsilon^3}d_p^2}}{{150{{(1 - \varepsilon )}^2}}},
\end{equation}

\begin{equation}
{F_\varepsilon } = \frac{{1.75}}{{\sqrt {150{\varepsilon ^3}} }},
\end{equation}
where $d_p$ is the solid particle diameter. To characterize the flow through porous media, two dimensionless numbers, namely the Reynolds number $Re$ and the Darcy number $Da$, are defined as

\begin{equation}
Re = \frac{{LU}}{\nu },
\end{equation}

\begin{equation}
Da = \frac{K}{{{L^2}}},
\end{equation}
where $L$ and $U$ are the characteristic length and velocity respectively. As the flow is fully developed between two plates, the velocity at the $x$ direction satisfies the following equation\cite{guo2002lattice}:

\begin{equation}
\label{e.velocity}
\left\{ \begin{split}
 &\nu\frac{{{\partial ^2}u}}{{\partial {x^2}}} - \frac{\varepsilon \nu }{K}u - \frac{{{\varepsilon}^2{F_\varepsilon }}}{{\sqrt K }}{u^2} = 0, \\
 &u(0) = 0, \quad u(H) = {u_0}. \\
 \end{split} \right.
\end{equation}

\begin{figure}
  \centering
  \includegraphics[width=6cm]{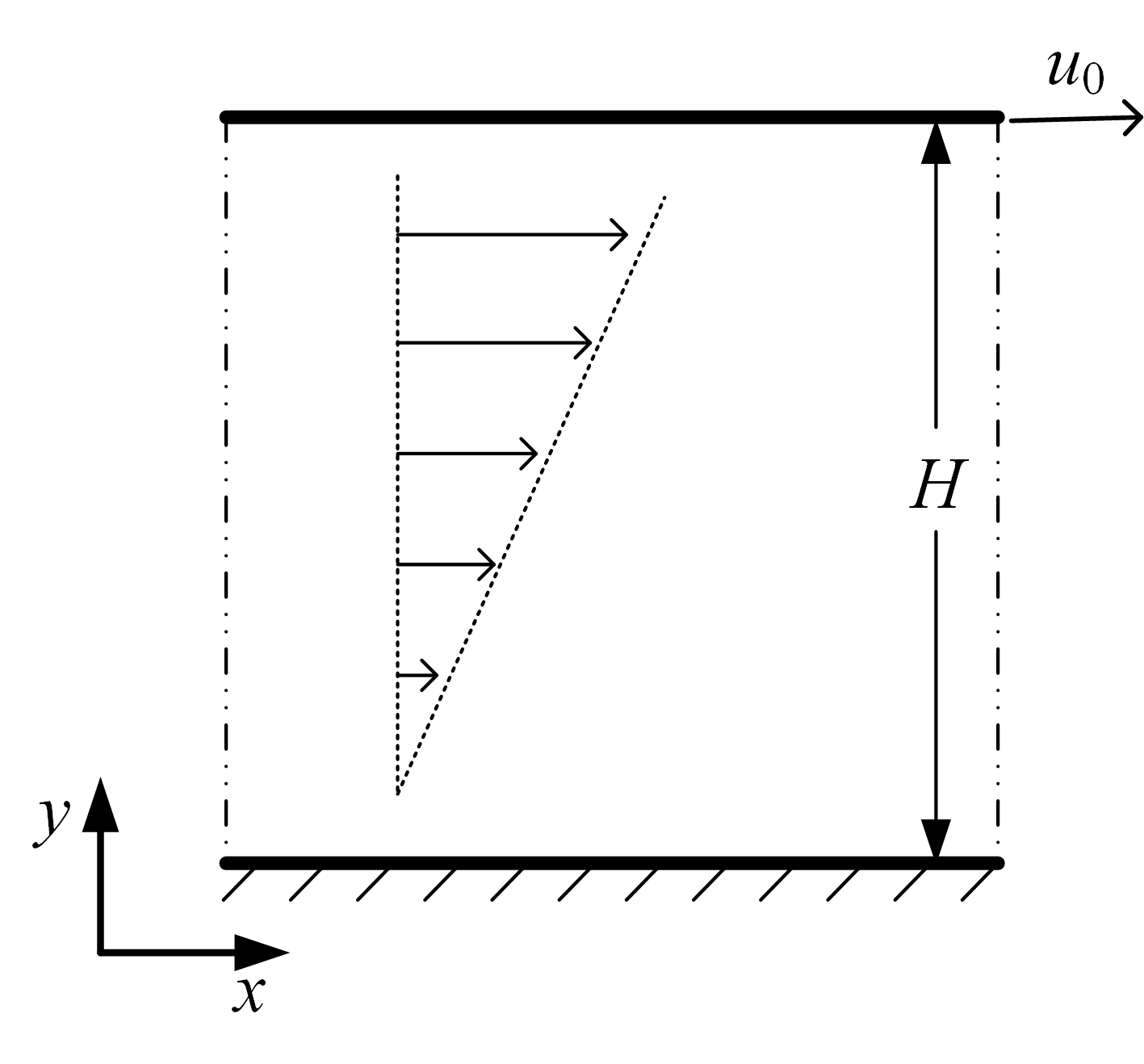}\\
  \caption{Illustration of the Couette flow through porous media.}
  \label{fig.1}
\end{figure}

In the simulations, the porous media are assumed as constant porosity $\varepsilon$ (chosen as 0.1), the characteristic velocity is set to be $u_0$ and the domain of porous media is divided into an $80\times80$ square lattice (with the characteristic length L is 80). When $Re=50$, the relaxation time $\tau$ is set as 0.6574; otherwise $\tau$ is set to be 0.8. The fluid density $\rho$ and fluid velocity $u$ are initialized as 1 and 0 respectively. The non-equilibrium extrapolation scheme\cite{guo2002boundary} is applied for both the no-slip boundary condition and velocity boundary condition. The solution shows that $u/U$ varies with $y/H$ at $x=H/2$ in fig.~\ref{fig.2}. The simulation results are found to be in excellent agreement with the reference solutions, proving that the present model can handle the Couette flow through porous media very well.

\begin{figure}
  \centering
  \subfigure[]{\includegraphics[width=7cm]{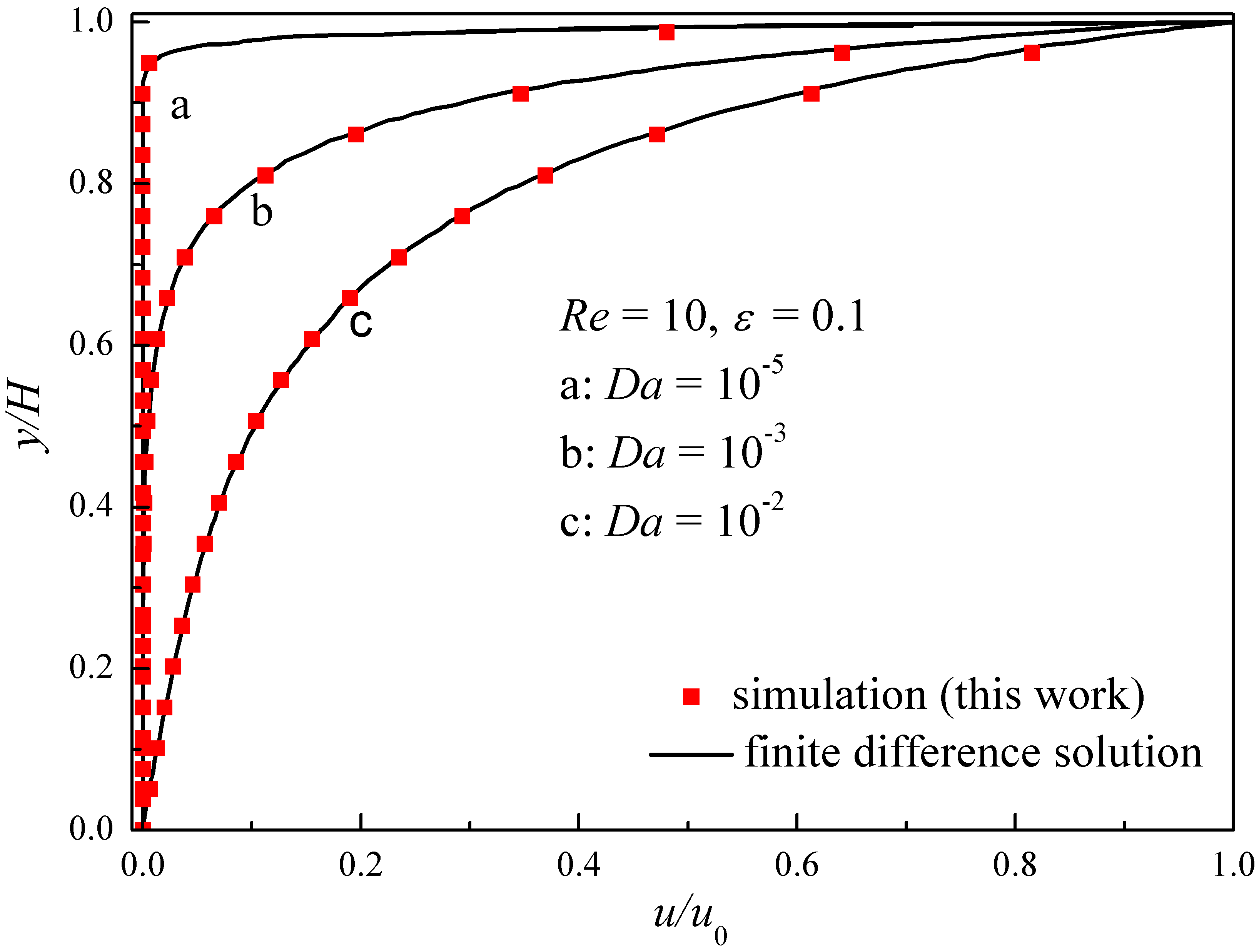}}\\
  \subfigure[]{\includegraphics[width=7cm]{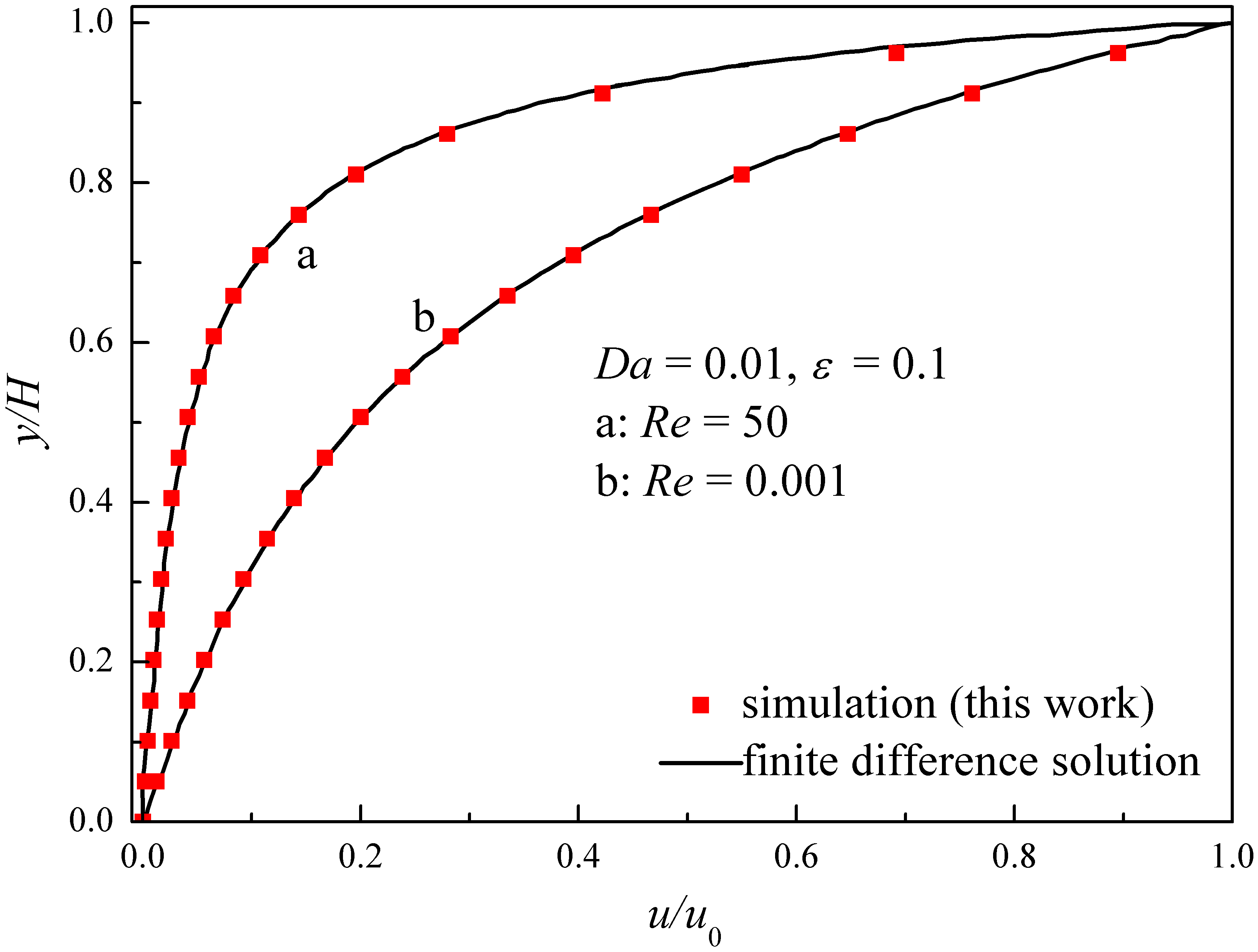}}
  \caption{Velocity profiles of the Couette flow along the $y$ axis at different $Re$ and $Da$. Symbols: the solution of the new model. Lines: finite difference solution of eq.~(\ref{e.velocity}), considered as the reference solution.}
  \label{fig.2}
\end{figure}

\section{Conclusions}
This letter is concerned with the lattice Boltzmann equation for solving the volume-averaged Navier-Stokes equations. Although most of previous numerical schemes have been carried for solving the volume-averaged Navier-Stokes equations, they are implicit/semi-implicit schemes. Here, we proposed a new LB model which is an explicit scheme for solving the equations. The proposed LB model provides an effective way in modeling porous flow and two-phase flow. We would like to point out that the volume fraction (or local porosity) $\varepsilon$ can be treated as a variable, which will be extended to solve the two-fluid model equations in future work.

\acknowledgments
The authors wish to thank Prof. Jinghai Li for his encouragement and help in improving the manuscript. The support from the National Natural Science Foundation of China (Grant No. 21106155) and the Chinese Academy of Sciences (Grant No. XDA07080303) is gratefully acknowledged.


\begin{thebibliography}{0}


\bibitem{HPP}
  \Name{Hardy J., Pomeau Y. \and De Pazzis O.}
  \REVIEW{Phys. Rev. Lett.}{31}{1973}{276}.

\bibitem{FHP}
  \Name{Frisch U., Hasslacher B.\and Pomeau Y.}
  \REVIEW{Phys. Rev. Lett.}{56}{1986}{1505}.

  \bibitem{Humieres1986}
  \Name{d'Humi\`{e}res D., Lallemand P.\and Frisch U.}
  \REVIEW{Europhys. Lett.}{2}{1986}{291}.

  \bibitem{Succi1989}
  \Name{Succi S., Foti E.\and Higuera F.}
  \REVIEW{Europhys. Lett.}{10}{1989}{433}.


  \bibitem{higuera1989a}
  \Name{Higuera F.J., Succi S. \and Benzi R.}
  \REVIEW{Europhys. Lett.}{9}{1989}{345}.

  \bibitem{higuera1989b}
  \Name{Higuera F.J. \and Jim\'{e}nez J.}
 \REVIEW{Europhys. Lett.}{9}{1989}{663}.

\bibitem{succi2001lattice}
  \Name{Succi S.}
  \Book{The lattice Boltzmann equation: for fluid dynamics and beyond}
  \Publ{Oxford university press}
  \Year{2001}.

\bibitem{he1997theory}
  \Name{He X.Y. \and Luo L.S.}
  \REVIEW{Phys. Rev. E}{56}{1997}{6811}.

  \bibitem{Koelman1991}
  \Name{Koelman J.M.V.A.}
  \REVIEW{Europhys. Lett.}{15}{1991}{603}.

  \bibitem{qian1992}
  \Name{Qian Y.H., d'Humi\`{e}res D. \and Lallemand P.}
  \REVIEW{Europhys. Lett.}{17}{1992}{479}.

  \bibitem{chen1998lattice}
  \Name{Chen S.Y. \and Doolen G.D.}
  \REVIEW{Annu. Rev. Fluid Mech.}{30}{1998}{329}.

  \bibitem{aidun2010lattice}
  \Name{Aidun C.K. \and Clausen J.R.}
  \REVIEW{Annu. Rev. Fluid Mech.}{42}{2010}{439}.

  \bibitem{qian1993}
  \Name{Qian Y.H. \and Orszag S.A.}
  \REVIEW{Europhys. Lett.}{21}{1993}{255}.

  \bibitem{sterling1996}
  \Name{Sterling J.D.\and Chen S.Y.}
  \REVIEW{J. Comp. Phys.}{123}{1996}{196}.

  \bibitem{He1997}
  \Name{He X.Y.\and Luo L.S.}
  \REVIEW{J. Stat. Phys.}{88}{1997}{927}.

  \bibitem{Qian1998}
  \Name{Qian Y.H.\and Zhou Y.}
  \REVIEW{Europhys. Lett.}{42}{1998}{359}.

  \bibitem{kuipers1992}
  \Name{Kuipers J.A.M., Van Duin K.J., Van Beckum F.P.H. \and Van Swaaij W.P.M.}
  \REVIEW{Chem. Eng. Sci.}{47}{1992}{1913}.

  \bibitem{VANS}
  \Name{Gidaspow D.}
  \Book{Multiphase Flow and Fluidization}
  \Publ{Academic Press}
  \Year{1994}.

  \bibitem{spalding1980numerical}
  \Name{Spalding D.B.}
  \Book{Numerical computation of multi-phase fluid flow and heat transfer}
  \Editor{Taylor C. \and Morgan K.}
  \Publ{Pinerdge}
  \Year{1980}
  \Page{139}.

  \bibitem{harlow1975numerical}
  \Name{Harlow F.H. \and Amsden A.A.}
  \REVIEW{J. Comput. Phys.}{17}{1975}{19}.

  \bibitem{karema1999efficiency}
  \Name{Karema H. \and Lo S.}
  \REVIEW{Compu. Fluids}{28}{1999}{323}.

  \bibitem{guo2002lattice}
  \Name{Guo Z.L. \and Zhao T.S.}
  \REVIEW{Phys. Rev. E}{66}{2002}{036304}.

  \bibitem{wang2005two}
  \Name{Wang T.F. \and Wang J.F.}
  \REVIEW{Phys. Rev. E}{71}{2005}{045301}.

  \bibitem{sankaranarayanan2008lattice}
  \Name{Sankaranarayanan K. \and Sundaresan S.}
  \REVIEW{Ind. Eng. Chem. Res.}{47}{2008}{9165}.

  \bibitem{sungkorn2012simulations}
  \Name{Sungkorn R. \and Derksen J.J.}
  \REVIEW{Phys. Fluids}{24}{2012}{123303}.

  \bibitem{eggels1995numerical}
  \Name{Eggels J.G.M. \and Somers J.A.}
  \REVIEW{Int. J. Heat Fluid Flow}{16}{1995}{357}.

  \bibitem{wang2013lattice}
  \Name{Wang L.M., Zhang B., Wang X.W., Ge W. \and Li J.H.}
  \REVIEW{Chem. Eng. Sci.}{101}{2013}{228}.

  \bibitem{force2002discrete}
  \Name{Guo Z.L., Zheng C.G. \and Shi B.C.}
  \REVIEW{Phys. Rev. E}{65}{2002}{046308}.

  \bibitem{wang2006cpc}
  \Name{Wang L.M., Ge W. \and Li J.H.}
  \REVIEW{Comput. Phys. Comm.}{174}{2006}{386}.

  \bibitem{zouhe1997boundary}
  \Name{Zou Q.S. \and He X.Y.}
  \REVIEW{Phys. of Fluids.}{6}{1997}{1591}.

  \bibitem{Ergun}
  \Name{Ergun S.}
  \REVIEW{Chem. Eng. Prog.}{48}{1952}{89}.

  \bibitem{vafai1984}
  \Name{Vafai K.}
  \REVIEW{J. Fluid. Mech.}{147}{1984}{233}.

  \bibitem{guo2002boundary}
  \Name{Guo Z.L., Zheng C.G. \and Shi B.C.}
  \REVIEW{Phys. Fluids}{14}{2002}{2007}.

\end{thebibliography}
\end{document}